\documentclass[12pt]{iopart}
\usepackage{iopams} 
\usepackage{graphicx}
\usepackage{braket}
\usepackage{threeparttable}
\usepackage{xspace}
\usepackage{bm}
\usepackage{hyperref}

\newcommand{\EF}{$E_\mathrm{F}$\xspace}
\newcommand{\Uf}{$\mathrm{U}~5f$\xspace}
\newcommand{\orb}[2]{$\mathrm{ #1 } ~ #2 $\xspace}
\newcommand{\hn}[1]{$h\nu #1~\mathrm{eV}$\xspace}

\newcommand{\pnt}[1]{$\mathrm{#1}$\xspace}
\newcommand{\Gm}{$\mathrm{\Gamma}$\xspace}
\newcommand{\lr}[1]{$[{#1}]$\xspace}

\newcommand{\URuSi}{$\mathrm{URu}_2\mathrm{Si}_2$\xspace}

\begin{document}

\title[{\small Electronic structure of $\mathrm{URu}_2\mathrm{Si}_2$  in paramagnetic phase: Three-dimensional ARPES study}]{Electronic structure of $\mathrm{URu}_2\mathrm{Si}_2$  in paramagnetic phase: Three-dimensional angle resolved photoelectron spectroscopy study}

\author{
Shin-ichi~Fujimori$^1$,
Yukiharu~Takeda$^1$, 
Hiroshi~Yamagami$^{1,2}$,
Etsuji~Yamamoto$^3$,
and Yoshinori~Haga$^3$
}

\address{$^1$ Materials Sciences Research Center, Japan Atomic Energy Agency, Sayo, Hyogo 679-5148, Japan}
\address{$^2$ Department of Physics, Faculty of Science, Kyoto Sangyo University, Kyoto 603-8555, Japan}
\address{$^3$ Advanced Science Research Center, Japan Atomic Energy Agency, Tokai, Ibaraki 319-1195, Japan}

\ead{fujimori@spring8.or.jp}
\vspace{10pt}
\begin{indented}
\item[] \today
\end{indented}

\begin{abstract}
The three-dimensional (3D) electronic structure of the hidden order compound $\mathrm{URu_2Si_2}$ in a paramagnetic phase was revealed using a 3D angle-resolved photoelectron spectroscopy where the electronic structure of the entire Brillouin zone is obtained by scanning both incident photon energy and detection angles of photoelectrons.  
The quasi-particle bands with enhanced contribution from the $\mathrm{U}~5f$ state were observed near $E_\mathrm{F}$, formed by the hybridization with the $\mathrm{Ru}~4d$ states.
The energy dispersion of the quasi-particle band is significantly depend on $k_z$, indicating that they inherently have a 3D nature.
The band-structure calculation qualitatively explain the characteristic features of the band structure and Fermi surface although the electron correlation effect strongly renormalizes the quasi-particle bands.
The 3D and strongly-correlated nature of the quasi-particle bands in $\mathrm{URu_2Si_2}$ is an essential ingredient for modeling its hidden-order transition.
\end{abstract}

\submitto{\href{https://iopscience.iop.org/journal/2516-1075/page/focus-on-electronic-structure-of-4f-and-5f-systems}{Focus on Electronic Structure of $4f$ and $5f$ Systems, Electronic Structure}}

\maketitle

\section{Introduction}
The hidden order transition in \URuSi at $T_\mathrm{HO}=17.5~\mathrm{K}$ \cite{URu2Si2_Palstra} has been an unresolved mystery in the condensed matter physics for more than 30 years.
Although numerous experimental and theoretical studies have been conducted on \URuSi, its order parameter has not been identified yet \cite{URu2Si2_review1,URu2Si2_review2,URu2Si2_review_2020}.
Angle-resolved photoelectron spectroscopy (ARPES) has also been applied to \URuSi to reveal its electronic structure \cite{URu2Si2_Itoh,Denlinger_XRu2Si2,URu2Si2_Andres1,URu2Si2_surf,URu2Si2_Yoshida1,URu2Si2_SXARPES,URu2Si2_trARPES,URuRh2Si2_SXARPES,URu2Si2_Yoshida2,URu2Si2_Boariu,URu2Si2_KShen,URu2Si2_Yoshida3,URu2Si2_Meng,URu2Si2_Cedric, ThRu2Si2_ARPES,URu2Si2_PNST,URu2Si2_Chen}, and some review papers exist \cite{URu2Si2_Tomasz_review,SF_review_JPCM,SF_review_JPSJ}.
Critical information from the ARPES studies is the \Uf state, which is the most essential information to establish an effective model for the hidden-order transition.
In all modern ARPES studies, quasi-particle bands with an enhanced \Uf contribution were observed near \EF, indicating that the crystal momentum $\bm{k}$ is a good quantum number for describing the \Uf states in \URuSi.
Furthermore, the experimental ARPES spectra were compared with the results of the band-structure calculation treating the \Uf states as an itinerant, and some correspondence between them was found \cite{Denlinger_XRu2Si2,URu2Si2_SXARPES,URu2Si2_Meng,URu2Si2_Cedric,ThRu2Si2_ARPES,URu2Si2_Chen}.
Therefore, the band-structure calculation treating the \Uf state as itinerant state captures an aspect of the \Uf state in \URuSi; however, the bands near \EF are strongly renormalized because of an electron correlation effect. 
Furthermore, band-structure calculations predict that the \Uf states are in the intermediate state with the $5f$ count of $n_f \sim 2.5$, consistent with the recent x-ray spectroscopy studies \cite{URu2Si2_EELS,URu2Si2_Booth,Ucore, URu2Si2_Compton}.
However, some recent x-ray spectroscopy studies proposed rather a localized \orb{U}{5f^2} ground state \cite{URu2Si2_Wray,URu2Si2_Kvashnina, URu2Si2_NIXS}, in line with the multipole scenario of the hidden order transition.
Thus, the \Uf state in \URuSi is still a controversial problem, which must be clarified. 

Furthermore, ARPES studies for \URuSi often reported different spectral profiles, which brought about confusion in their coherent understanding \cite{SF_review_JPCM}.
Table~\ref{URu2Si2_ARPES} summarizes the ARPES study of \URuSi that was conducted in different experimental setups, such as different photon energies and sample temperatures.
The cause of the confusion is mostly because of the different photon energies used in these studies, resulting in different experimental conditions, such as different out-of-plane momenta ($k_\perp$), surface sensitivities ($\lambda$), and matrix element effects ($|M_{i,f}|^2$).
Specifically, the out-of-plane momentum $k_\perp$ was not clearly defined in many ARPES studies for \URuSi\cite{URu2Si2_Itoh,URu2Si2_Andres1,URu2Si2_Yoshida1,URu2Si2_trARPES,URu2Si2_Yoshida2,URu2Si2_KShen,URu2Si2_Chen}, and the three-dimensional (3D) nature of its electronic structure was not considered.
Meng \etal \cite{URu2Si2_Meng} explored this point by mapping the out-of-plane Fermi surface map with $h\nu=14$--$34~\mathrm{eV}$.
However, ARPES experiments with low photon energies have enhanced the surface sensitivity of $\lambda \lesssim 5~\mathrm{\AA}$, and the surface electronic structure is dominant in their spectra.
Furthermore, in their experiment, the $k_\perp$ broadening is estimated to be $\Delta k_\perp = 1/\lambda \sim 0.2~\mathrm{\AA}^{-1}$, which is almost half the size of the Brillouin zone along the $k_\perp$ direction ($2k_c \sim 0.45~\mathrm{\AA}^{-1}$).
Chen \etal highlighted another surface-related issue and reported two types of surface termination in \URuSi, resulting in completely different ARPES spectral profiles.

In this study, we reveal the complete steric electronic structure of \URuSi in the paramagnetic phase using the 3D scan of soft X-ray (SX) ARPES spectra.
The SX-ARPES is superior in its enhanced bulk sensitivity and good $k_\perp$ resolution, enabling us to attend to the problems in ARPES experiments \cite{SF_review_JPSJ}.
In our previous SX-ARPES studies for \URuSi \cite{URu2Si2_SXARPES,URuRh2Si2_SXARPES,ThRu2Si2_ARPES,URu2Si2_PNST,SF_review_JPSJ}, the \Uf states of \URuSi formed quasi-particle bands near \EF.
However, the ARPES scans were done within limited high-symmetry planes, and the electronic structure's three-dimensionality was not clarified.
Furthermore, several high-energy-resolution ARPES studies have been reported, and the relationship between our results and these studies have been not clarified.
We elucidated the steric electronic structure of \URuSi in the paramagnetic phase, and compared the results with the band-structure calculation results.
The detailed electronic structure of symmetry points are discussed, and the relationship to previous ARPES studies are also considered.

%-----------------------------------------------------------------------------------------
\begin{table*}[tb]
\caption{Summary of ARPES studies of \URuSi.}
	\begin{tabular}{lllllll}
		\hline\hline
		 year	& light source & $h\nu$ (eV) & $\Delta E$ (meV)	& $T_\mathrm{sample}$ (K)	& phase(s)	& reference	\\
		\hline
		1999	& He I					& 21.2			& 30--50			& 30			& PM			& \cite{URu2Si2_Itoh}			\\
		2001	& ALS	BL7.0			& 85-156		& 60				& $<150$		& PM			& \cite{Denlinger_XRu2Si2}	\\
		2009	& He I					& 21.2			& 5				& 10--26		& PM + HO		& \cite{URu2Si2_Andres1} 	\\
		2010	& Laser				& 7			& 2				&  7--25		& PM + HO		& \cite{URu2Si2_Yoshida1} 	\\
		2011	& SPring-8 BL23SU	& 760--800	& 140				&  20			& PM			& \cite{URu2Si2_SXARPES} 	\\
				& SRC 71A$^\ast$	& $\sim 30$	& 15				& 12--19		& PM + HO		& \cite{URu2Si2_trARPES} 	\\
		2012	& Laser				& 7			& 1.5				& 5--21		& PM + HO		& \cite{URu2Si2_Yoshida2} 	\\
		2013	& BESSY II			& 8.4--31		& 3				& 1--68		& PM + HO		& \cite{URu2Si2_KShen} 		\\
				& BESSY II			& 21--69		& $\lesssim 7$	& 2--20		& PM + HO		& \cite{URu2Si2_Boariu} 		\\
				& SRC					& 14--98		& 15--20			& 8--35		& PM + HO		& \cite{URu2Si2_Meng} 		\\
				& BESSY II			& 19--51		& 4--7				& 2--22		& PM + HO		& \cite{URu2Si2_Yoshida3} 	\\
		2014	& BESSY II			& 21--50		& 4--7				& 1--20		& PM + HO		& \cite{URu2Si2_Cedric} 		\\
		2018	& He I					& 21.2			& $\lesssim 15$	& 11--82		& PM + HO		& \cite{URu2Si2_Chen} 		\\
		\hline\hline
	\end{tabular}
		$^\ast$time resolved ARPES
	\label{URu2Si2_ARPES}
\end{table*}
%-----------------------------------------------------------------------------------------

\section{Experimental Procedure}
Photoemission experiments were conducted at the SX beamline SPring-8 BL23SU \cite{BL23SU2}.
High-quality single crystals were grown using the Czochralski method.
Clean sample surfaces were obtained by cleaving the samples {\it in situ} perpendicular to the $c$-axis under an ultra-high vacuum (UHV) condition.
The sample temperature was kept at $20~\mathrm{K}$ during the course of measurements.
The VG Scienta SES-2002 electron analyzer was used for ARPES measurements, and the overall energy resolution was about $\sim 100~\mathrm{meV}$.
%-----------------------------------------------------------------------------------------
\begin{figure*}
	\centering
	\includegraphics[scale=0.45]{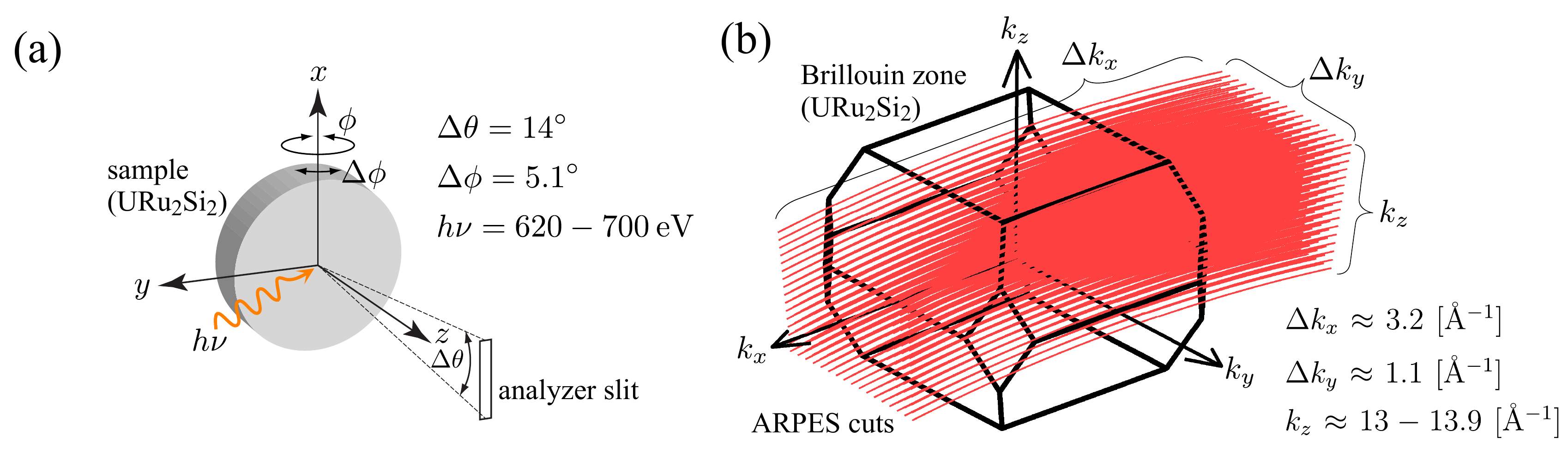}
	\caption{Experimental configuration and the ARPES cuts in the momentum space.
	(a) The experimental setup (b) ARPES cuts and Brillouin zone of \URuSi in the paramagnetic phase.
}
	\label{3DBZ}
\end{figure*}
%-----------------------------------------------------------------------------------------
Figure~\ref{3DBZ} summarizes our experimental setup and ARPES cuts in the momentum space.
In our experimental setup (Fig.~\ref{3DBZ}(a)), the scan along the slit direction of the electron analyzer is defined as $\theta$.
The angular window along the $\theta$ direction is $\Delta \theta \approx 14^\circ$.
The electron analyzer's two-dimensional (2D) detector obtains the ARPES spectra along the $\theta$ direction.
The sample was rotated perpendicular to the $\phi$ direction (Fig.~\ref{3DBZ} (a)).
The scan along the $\phi$ direction was made by the angular range of $\Delta \phi = 5.1^\circ$ with an angular step of $0.3^\circ$.
Along the photon energy direction, the ARPES scans by $h\nu=620$--$700~\mathrm{eV}$ with every $5~\mathrm{eV}$ step were performed.
In total, the one-dimensional $\theta$ scans were $17 \times 18 = 306$.
The ARPES cut positions were determined by assuming a free-electron final state, where the electron momentum is expressed as

\begin{eqnarray}
	k_x  & = &  \frac{ \sqrt{2 m E_\mathrm{kin}}}{\hbar} \sin{\theta} \cos{\phi}, \nonumber\\
	k_y  & = &  \frac{ \sqrt{2 m E_\mathrm{kin}}}{\hbar} \sin{\phi} - k_{\parallel \mathrm{photon}}, \nonumber\\
	k_z  & = & \sqrt{ \frac{2 m}{\hbar^2}(E_\mathrm{kin} \cos^2{\theta} \cos^2{\phi} + V_0) } - k_{\perp \mathrm{photon}}
	\label{kxkykz} 
\end{eqnarray}

Because the incident photon's direction was within the $yz$ plane in our experimental setup, the photon momentum	 does not contribute to the $k_x$ direction.
We took the inner potential as $V_0=12~\mathrm{eV}$, determined by our previous ARPES for \URuSi \cite{URu2Si2_SXARPES}.
From Eq.~\ref{kxkykz}, the ARPES cuts of the 3D scans cover the cuboid volume with $\Delta k_x \approx 3.2~\mathrm{\AA ^{-1}}$, $\Delta k_y \approx 1.1~\mathrm{\AA ^{-1}}$, and $k_z \approx 13 - 13.9~\mathrm{\AA ^{-1}}$ in the momentum space (Fig.~\ref{3DBZ} (b)).
The figure also shows the Brillouin zone of \URuSi in the paramagnetic phase.
The distances from the \Gm point to the boundary of the Brillouin zone are $\sim 0.90~\mathrm{\AA ^{-1}}$ and $0.66~\mathrm{\AA ^{-1}}$ along the $k_x$ and $k_z$ directions, respectively, and our ARPES cuts cover more than a quarter of the Brillouin zone. 

The obtained spectra were normalized to the intensity of the incident photons, estimated from the dark intensity of each spectrum, namely, the spectral intensity above the \EF.
The obtained data were four-fold symmetrized according to the the crystal structure's symmetry to eliminate the influence from the matrix element effect originated from the asymmetries in this experimental setup.
During the measurements, the vacuum was typically $<1 \times 10^{-8}~\mathrm{Pa}$, and the sample surfaces were stable during the measurements (3--4 days) because no significant changes were recognized in the ARPES spectra during the period.
Background contributions in ARPES spectra originated from elastically scattered photoelectrons because of surface disorder or phonons were subtracted by assuming the momentum-independent spectrum, whose procedure is described in Ref.~\cite{UGe2_UCoGe_ARPES}.

\section{Results and Discussion}
%-----------------------------------------------------------------------------------------
\subsection{3D Fermi surface of \URuSi }
\begin{figure*}
	\centering
	\includegraphics[scale=0.45]{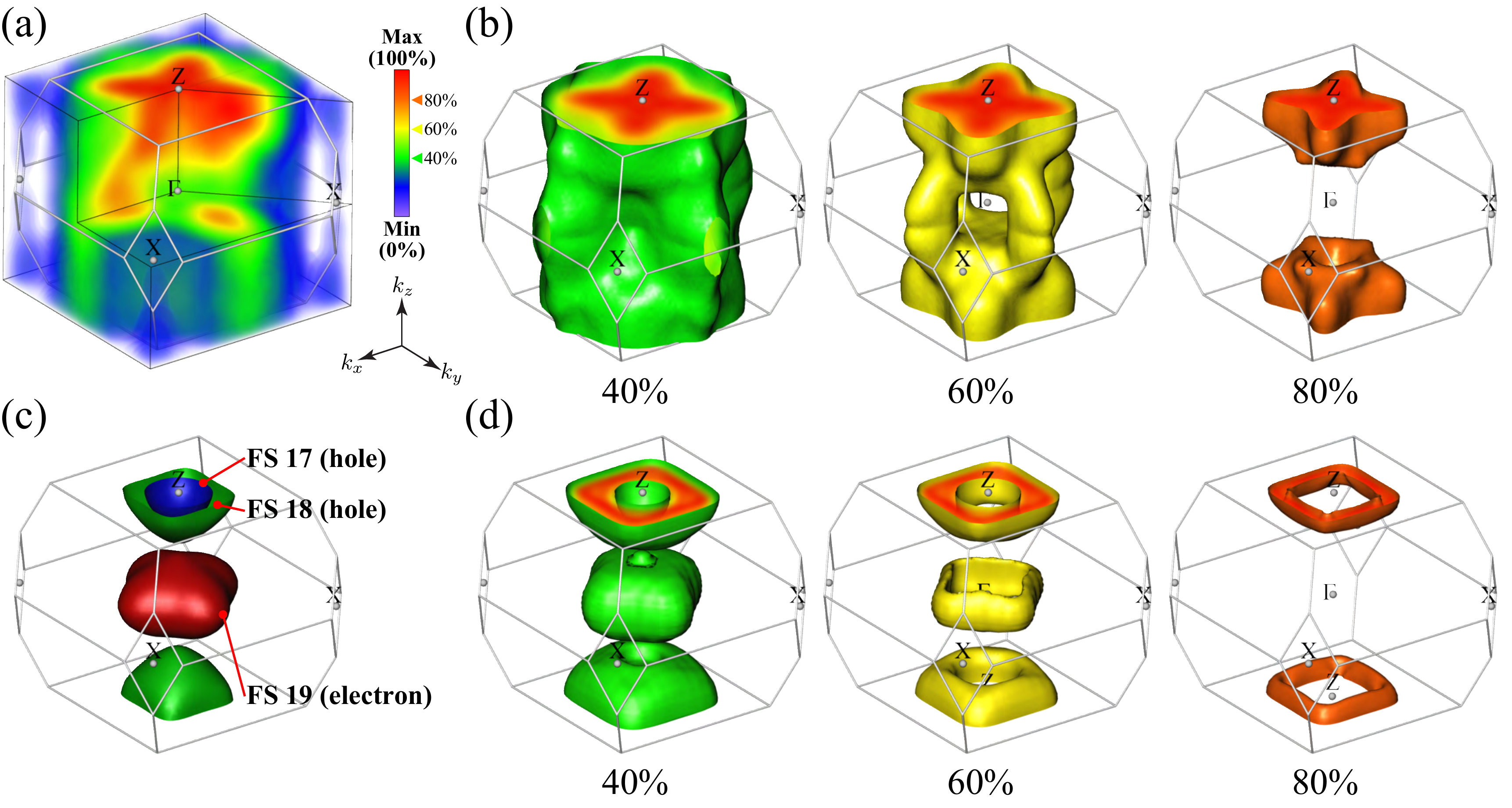}
	\caption{3D volume Fermi surface mapping and the result of the band-structure calculation.
	(a) Experimental 3D volume Fermi-surface mapping of \URuSi.
	The minimum and maximum intensities were defined as 0\% and 100\%, respectively.
	(b) Isosurfaces of 3D volume Fermi-surface mapping with the intensities of 40\%, 60\%, and 80\%.
	(c) Calculated Fermi surface of \URuSi.
	(d) Simulated 3D volume Fermi-surface mapping based on the band-structure calculation.
}
	\label{3DFS}
\end{figure*}
%-----------------------------------------------------------------------------------------
First, we present the 3D Fermi surface map of \URuSi to show the 3D nature of the Fermi surface.
The contributions from the \Uf and \orb{Rh}{4d} states are dominant in this photon energy range \cite{Atomic}, but the states near \EF predominantly consist of the contribution from the \Uf states \cite{URu2Si2_SXARPES}.
Thus, the Fermi surface maps substantially represent contributions from the \Uf states.
Figure~\ref{3DFS} shows the 3D distribution of the ARPES intensity integrated over $E_\mathrm{F} \pm 50~\mathrm{meV}$ and its comparison with the band-structure calculation results.
Figure~\ref{3DFS} (a) shows the 3D Fermi surface map's rendering.
Hereafter, we designate the minimum and maximum photoemission intensities within the entire Brillouin zone as 0\% and 100\%, respectively.
The 3D Fermi surface map has complex 3D distributions, demonstrating that \URuSi has an inherent 3D electronic structure.
Overall, a cylindrical region exists, with an enhanced intensity along the $k_z$ direction, whereas the intensities in the in-plane corners of the Brillouin zone is rather weak.
To see the details of the Fermi surface map's 3D structure, the isosurfaces of the ARPES intensities with 40\%, 60\%, and 80\% are shown in Fig.~\ref{3DFS} (b). 
The isosurface with an intermediate intensity of 40\% has 2D cylindrical structure, but those of higher intensities (60\% and 80\%) have more complex 3D distributions.
There exists a cage-like structure around the \Gm point in the isosurface with 60\% intensity and a blob-like feature with an enhanced intensity around the \pnt{Z} point in the isosurface with 80\% intensity.
Thus, the ARPES intensity is more enhanced around the \pnt{Z} point than around the \Gm point, consistent with previous ARPES studies on \URuSi  \cite{URu2Si2_SXARPES,URu2Si2_Yoshida2,URu2Si2_Yoshida3,URu2Si2_Meng,URu2Si2_Cedric}.
Another interesting point is that the in-plane intensity distribution is different between in the \Gm and \pnt{Z} planes; the intensity is enhanced along the \lr{100} and equivalent directions around the \Gm point, whereas it is enhanced along the \lr{110} and equivalent directions around the \pnt{Z} point.

We compare the experimental data with the band-structure calculation's results, treating the \Uf states as an itinerant.
The calculation is identical to that in Ref.\cite{URu2Si2_SXARPES}.
Figure~\ref{3DFS} (c) shows the calculated Fermi surfaces.
Two hole-type Fermi surfaces 17 and 18 around the \pnt{Z} point and one electron-type Fermi surface 19 around the \Gm point  exist in the band-structure calculation.
Figure~\ref{3DFS} (d) shows the isosurfaces of the simulated ARPES intensity based on the band-structure calculation, which can be directly compared with the experimental results shown in Fig.~\ref{3DFS} (b).
This simulation considered photoionization cross-section of atomic orbitals and experimental energy and momentum resolutions.
Ref.~\cite{UN_ARPES} describes the simulation details.
There exist some similarities between the experimental results and the simulation.
For example, the intensities are more enhanced around the \pnt{Z} point than around the \Gm point.
Furthermore, the in-plane anisotropies of the intensity distribution between the experimental data and calculation results correlate.
The intensity is more enhanced along the \lr{100} and \lr{110} directions around the \Gm point and \pnt{Z} point, respectively.
However, the structures in the experimental map are much broader than the calculation.
Note that the in-plane intensity around the \Gm point appears only along the \lr{100} and the equivalent directions, and the intensities between them almost disappeared because of the matrix element effect, as discussed in our previous SX-ARPES study \cite{URu2Si2_SXARPES}.
Accordingly, the comparison between experimental and the calculated Fermi surface maps indicate that there exist some similarities between the experimental and calculated Fermi surface maps, but there are some quantitative disagreements between them.
\subsection{3D band structure of \URuSi}
%-----------------------------------------------------------------------------------------
\begin{figure*}
	\centering
	\includegraphics[scale=0.45]{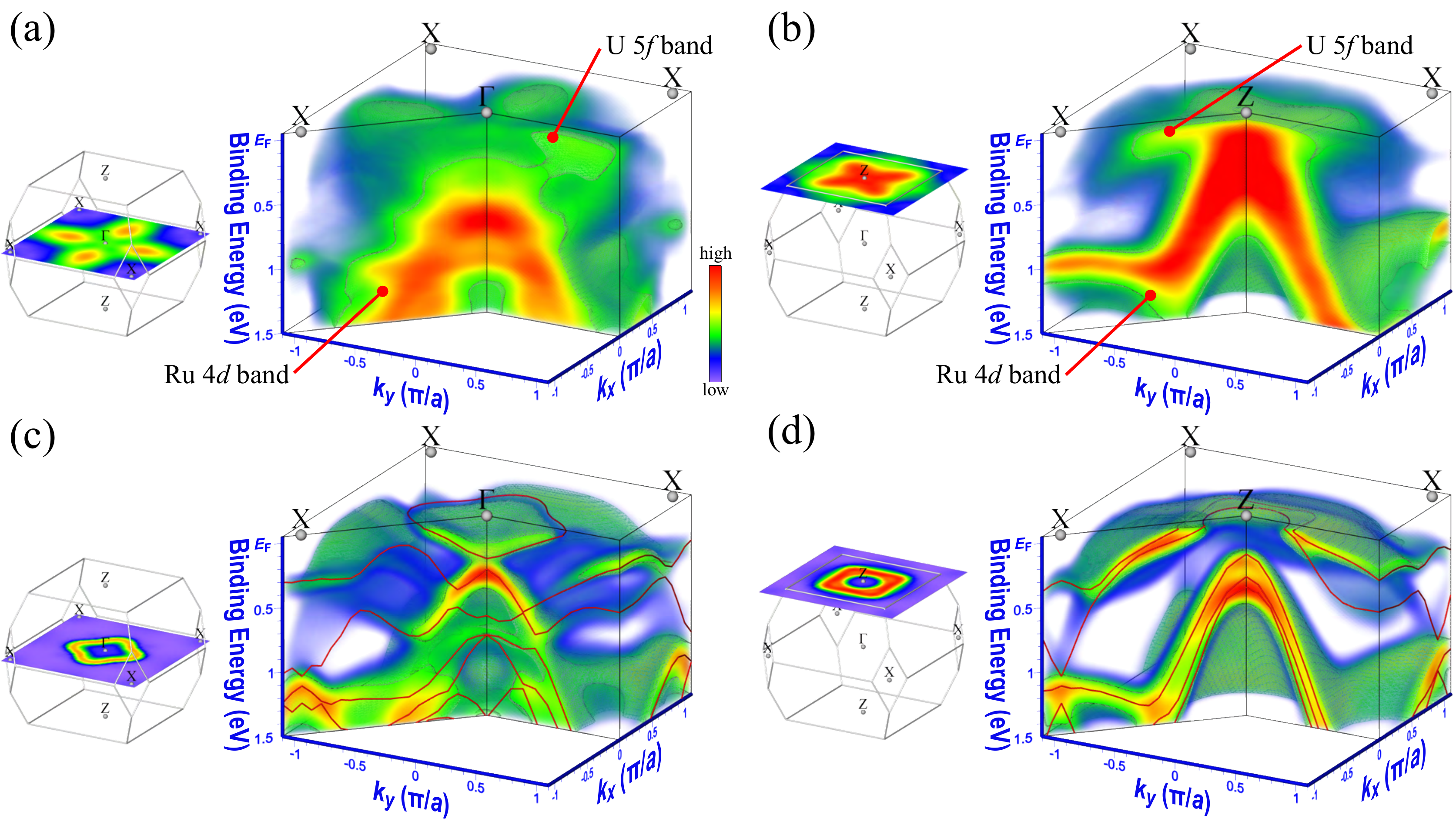}
	\caption{Experimental ARPES spectra of \URuSi within the $k_x$--$k_z$ planes and their simulations based on band-structure calculations.
The horizontal and vertical axes correspond to the momentum ($k_x$ and $k_y$) and binding energy, respectively.
The 2D Fermi surface maps with corresponding planes in the Brillouin zone are also presented.
	(a,b) Experimental 3D band structure (c,d) the simulation results based on the band-structure calculation.
}
	\label{3Dbandkxky}
\end{figure*}
%-----------------------------------------------------------------------------------------
Figure~\ref{3Dbandkxky} shows the experimental ARPES spectra within the $k_x$--$k_y$ planes and their simulations based on the band-structure calculation. 
The horizontal and vertical axes are the crystal momenta ($k_x$ and $k_y$) and binding energy ($E_\mathrm{B}$), respectively.
Figures~\ref{3Dbandkxky} (a) and (b) shows the ARPES spectra's volume renderings within the \Gm ($k_z=40\pi/c$) and \pnt{Z} ($k_z=42\pi/c$) planes, respectively.
The most prominent features in these spectra are inverted cone-like dispersions centered at the \Gm or \pnt{Z} points, predominantly contributed from the \orb{Ru}{4d} states.
There exist relatively flat features near \EF, attributed to the quasi-particle bands with dominant contributions from the \Uf states.
The quasi-particle bands' intensity around the \pnt{Z} point is enhanced at the apex of the \orb{Ru}{4d} bands, demonstrating that the \Uf states are strongly hybridized with the \orb{Ru}{4d} states.
Around the \Gm point, there exist blob-like features near \EF along the \lr{100} and equivalent directions.
According to detailed analysis \cite{URu2Si2_SXARPES,ThRu2Si2_ARPES,SF_review_JPSJ}, this feature corresponds to the electron Fermi surface around the \Gm point and the hole Fermi surface around the \pnt{Z} point.
Note that the features are observed only along the \lr{100} and equivalent directions and disappear for other directions because of the matrix element effect, as discussed in the Fermi surface map shown in Fig.~\ref{3DFS}.

Figures~\ref{3Dbandkxky} (c) and (d) show the the simulation results based on the band-structure calculation.
Overall, the calculated band structure has a much more complicated structure than that of experimental spectra, but some qualitative agreements can be found between the experimental spectra and the band-structure calculation results.
For example, the experimental \orb{Ru}{4d} bands have some correspondence to the calculated dispersions.
Because the \Uf--\orb{Ru}{4d} hybridization strongly influences the \orb{Ru}{4d} bands' structure \cite{ThRu2Si2_ARPES}, the good correlation indicates that \Uf states contribute to the band structure of \URuSi.
Furthermore, the overall quasi-particle band's energy dispersions within the \pnt{Z} plane have similar hole-like dispersions in the experiment and calculation.
However, the states near \EF are less dispersive than the calculation results.
For example, the hole-like dispersion around the \pnt{Z} point in the experiment is less dispersive and more featureless than the calculation results.
Such discrepancies between the experiment and calculation are recognized specifically near \EF, and thus they originate from the renormalization of the \Uf bands by the strong correlation effect.

%-----------------------------------------------------------------------------------------
\begin{figure*}
	\centering
	\includegraphics[scale=0.45]{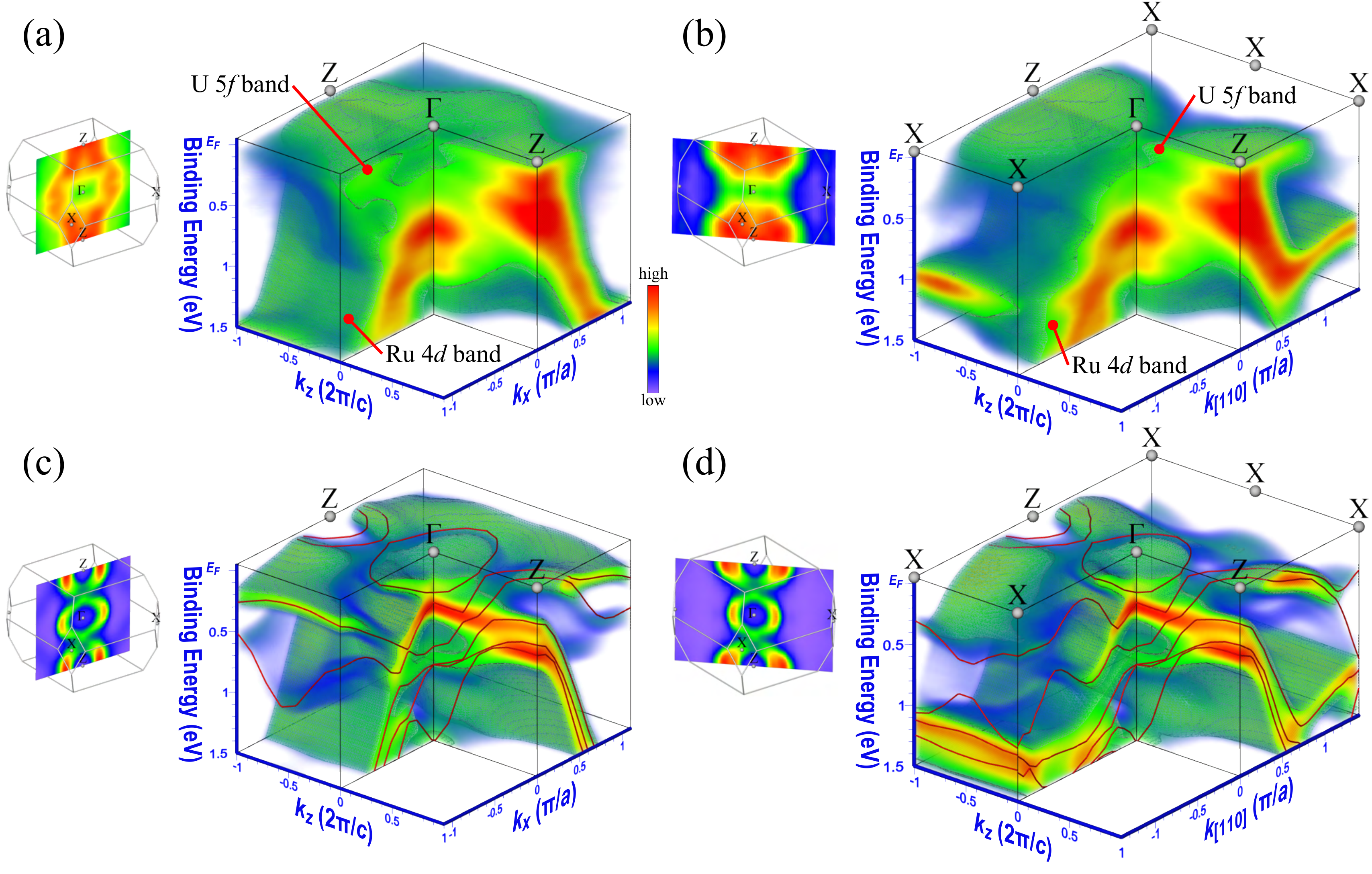}
	\caption{
Experimental ARPES spectra of \URuSi within the ($k_x$, $k_y$)--$k_z$ planes and their simulations based on the band-structure calculations.
The horizontal and vertical axes correspond to the momentum ($k_x$, $k_{[110]}$, or $k_z$) and binding energy, respectively.
	(a,b) Experimental 3D band structure (c,d) the simulation results based on the band-structure calculation.
}
	\label{3Dbandkxkz}
\end{figure*}
%-----------------------------------------------------------------------------------------
To further reveal the 3D nature in the electronic structure of \URuSi, Fig.~\ref{3Dbandkxkz} shows the ARPES spectra along the $k_z$ direction and corresponding Fermi surface maps.
Figures~\ref{3Dbandkxkz} (a) and (b) show the ARPES spectra within the $k_x$--$k_z$ and $k_{[110]}$--$k_z$ planes, respectively.
The overall structure of the energy dispersions is similar to the case of the in-plane spectra shown in Fig.~\ref{3Dbandkxky}.
The prominent inverted parabolic dispersions along the $k_x$ or $k_{[110]}$ directions over the entire binding energies are the contributions predominantly from the \orb{Ru}{4d} states.
The \orb{Ru}{4d} bands are less dispersive along the $k_z$ direction, indicating that the quasi-2D nature of the electronic structure in $\mathrm{Ru_2 Si_2}$ layers.
The distribution of the narrow \Uf bands near \EF has distinct dependency on $k_z$. 
The intensity of the \Uf bands is enhanced around the \pnt{Z} point, where the \orb{Ru}{4d} bands approach the \EF.
This again indicates that the \Uf states are hybridized with the \orb{Ru}{4d} bands around the \pnt{Z} point and the \Uf -- \orb{Ru}{4d} hybridization forms the heavy quasi-particle $5f$ bands.
There is a doughnut-like feature near \EF around the \Gm point, which is also formed by the quasi-particle \Uf bands.
The feature is only observed along the $k_x$ and equivalent directions, because of the matrix element effect, as discussed in the Fermi surface map shown in Figs.~\ref{3DFS} (a) and (b).

Figures~\ref{3Dbandkxkz} (c) and (d) show the simulation of the ARPES spectra based on the band-structure calculation, corresponding to experimental spectra shown in Figs.~\ref{3Dbandkxkz} (a) and (b), respectively.
The \orb{Ru}{4d} bands are less dispersive along the $k_z$ direction, whose structure is similar to the experimental band dispersion.
However, the correlation between experiment and calculation is not obvious near \EF, where the \Uf states' contribution is enhanced.
For example, there exist the electron-like
For example, the appearance of the electron-like band around the \Gm point shown in Fig.~\ref{3Dbandkxkz} (c) is different from that of the experimental band shown in Fig.~\ref{3Dbandkxkz} (a) although the Fermi surface maps shown in the left of each ARPES spectra are similar each other.
The features observed at the \pnt{Z} point are similar between the experiment and calculation, but some differences exist in their details.
The experimental energy dispersions are more featureless and less dispersive than in the calculation.
This again indicates that the experimental \Uf bands are strongly renormalized, consistent with the much more enhanced specific heat coefficient in the experiment than in the band-structure calculation.
Thus, the correlation between the experimental data and calculation results is similar to the case of the in-plain band structure:
The experimental \orb{Ru}{4d} band structure corresponds well with the calculation results but the \Uf bands are more featureless than the calculation originating from the strong electron correlation effect.
Nevertheless, the \Uf states strongly depend on $k_z$, indicating that the three dimensionality is essential in the electronic structure of \URuSi.

\subsection{Discussion}
Accordingly, the nature of the \Uf bands in \URuSi was revealed by observing the entire Brillouin zone by 3D-ARPES. 
The \Uf state in \URuSi forms 3D quasi-particle bands by hybridization with the electronic states in $\mathrm{Ru_2Si_2}$ layers with a quasi-2D nature.
The result argues that the three-dimensionality of \Uf states is a critical aspect in the electronic structure of \URuSi.
Furthermore, the 3D Fermi surface and the overall band structure have some correspondence between the experiment and the band-structure calculation treating \Uf states as an itinerant.
Here, we further discuss the relationship between this result and other ARPES studies at each high-symmetry points.

%-----------------------------------------------------------------------------------------
\paragraph{\Gm point}
%-----------------------------------------------------------------------------------------
Around the \Gm point, a cage-like feature in the 3D Fermi surface map exists (Fig.~\ref{3DFS} (b)), which is a fragment of the electron-type Fermi surface, as observed in the energy dispersions shown in Fig~\ref{3Dbandkxky} (a).
Similar band dispersions were also observed in the ARPES spectra measured at $h\nu=156~\mathrm{eV}$ \cite{Denlinger_XRu2Si2}.
However, ARPES studies with low photon energies of \hn{\lesssim 50} reported two rapid hole-like dispersions around the \Gm point \cite{URu2Si2_surf,URu2Si2_Boariu,URu2Si2_KShen,URu2Si2_Yoshida3}, which are absent in this study.
One of the hole-like dispersions with its apex at $E_\mathrm{B} \sim 30~\mathrm{meV}$ was identified as the surface electronic structure because its energy position has a time dependence \cite{URu2Si2_surf}.
Another hole-like dispersion forming the Fermi surface at $k_\mathrm{F} \sim \pm 0.15~\mathrm{\AA^{-1}}$ was not clearly observed in the ARPES study with \hn{=50} \cite{URu2Si2_Cedric}, although the spectra also correspond to the scan around the \Gm point.
This indicates that the band originates from the surface electronic structure because its existence is sensitive to the cleaved surface.
Furthermore, similar energy dispersions were recognized in ARPES experiments with low photon energies of \hn{=7-51} \cite{URu2Si2_Andres1,URu2Si2_surf,URu2Si2_Yoshida1,URu2Si2_trARPES,URu2Si2_Yoshida2,URu2Si2_Boariu,URu2Si2_KShen,URu2Si2_Yoshida3,URu2Si2_Cedric,URu2Si2_Chen}, namely, any $k_z$ with surface sensitive photon energies.
Thus, the band originates from the 2D surface state.
By considering this result and the SX-ARPES \cite{URu2Si2_SXARPES}, there should be an electron-like Fermi surface around the \Gm point whose shape would correspond to the one in the band-structure calculation, though it is strongly renormalized near \EF. 
%-----------------------------------------------------------------------------------------
\paragraph{\pnt{Z} point}
%-----------------------------------------------------------------------------------------
In this study, the hole-type energy dispersion with enhanced contributions from the \Uf states were observed around the \pnt{Z} point.
The feature corresponds to the two hole-type Fermi surfaces, as observed in previous SX-ARPES studies \cite{URu2Si2_SXARPES}, whose topologies are consistent with the band-structure calculation results, even though their sizes are larger in the experiment than in the calculation.
In low-$h\nu$ ARPES studies, an enhanced intensity exists around the \pnt{Z} point, consistent with this result \cite{Denlinger_XRu2Si2,URu2Si2_Boariu,URu2Si2_KShen,URu2Si2_Yoshida3,URu2Si2_Meng,URu2Si2_Cedric}.
However, the hole-type energy dispersion with smaller size was observed in low-$h\nu$ ARPES studies \cite{URu2Si2_Andres1,URu2Si2_surf,URu2Si2_Yoshida1,URu2Si2_trARPES,URu2Si2_Yoshida2,URu2Si2_Boariu,URu2Si2_KShen,URu2Si2_Yoshida3,URu2Si2_Cedric,URu2Si2_Chen}, 
 and no corresponding band exists in the band-structure calculation.
Because its shape is similar to that of the hole-type dispersion observed in the low-$h\nu$ ARPES spectra at the \Gm point, it also originates from the surface electronic structure.
%-----------------------------------------------------------------------------------------
\paragraph{\pnt{X} point}
%-----------------------------------------------------------------------------------------
As discussed in Ref.~\cite{SF_review_JPCM}, some confusion exist about the electronic structure around the \pnt{X} point.
The low-$h\nu$ ARPES studies of \URuSi reported the Fermi surface around the \pnt{X} point \cite{Denlinger_XRu2Si2,URu2Si2_Boariu,URu2Si2_Chen}.
Recently, Chen \etal reported that the band dispersion is different, depending on the types of surface terminations  \cite{URu2Si2_Chen}.
Electron-like and hole-like Fermi surfaces were observed in the $\mathrm{U}$-- and $\mathrm{Si}$--terminated surfaces, respectively.
Temperature dependence of the hole-like band at the \pnt{X} point was also reported \cite{Denlinger_XRu2Si2,URu2Si2_Boariu,URu2Si2_Chen}, where the formation of heavy quasi-particle bands was observed at low temperatures.
However, some band-structure calculations predicted shallow electron-like Fermi surfaces around the \pnt{X} point \cite{Denlinger_XRu2Si2,URu2Si2_nmat,URu2Si2_Oppeneer,URu2Si2_Ikeda}, but their band structures are different from the mentioned experimental observation.
In the experimental spectra, there exist energy dispersions with an order of $1~\mathrm{eV}$, but the bottom of the electron-type band in the band-structure calculation is at $< 10~\mathrm{meV}$.
Thus, the electronic structure at the \pnt{X} point differs in low-$h\nu$ ARPES and the calculations, though both may predict a similar- shaped Fermi surface around the \pnt{X} point.
In this study, there exist weak intensities around the \pnt{X} point (Fig.~\ref{3DFS} (a))), which have a 2D behavior.
Thus, the state at the \pnt{X} point reported in previous ARPES studies is the contribution from the surface electronic structure, and no Fermi surface exists at the \pnt{X} point originated from the bulk electronic structure.
\\ \\
Finally, we comment on the relationship between this result and the mechanism of the hidden-order transition.
Although this study could not reveal the detailed topology of the Fermi surface of \URuSi, some similarities were found between the experimentally-observed band structures and Fermi surfaces and the band-structure calculation.
Thus, the itinerant nature of \Uf states should be an essential aspect in understanding the order parameter of the hidden-order transition.
The 3D nature of the quasi-particle bands in \URuSi further argues that the three-dimensionality in the electronic structure of \Uf states must be considered in modeling the hidden-order transition. 

\section{Conclusion}
We applied the 3D-ARPES to the hidden-order compound \URuSi and revealed its detailed electronic structure in the paramagnetic phase.
The \Uf bands form the quasi-particle bands with an inherently 3D nature by hybridization with the electronic states in the $\mathrm{Ru_2 Si_2}$ layers with quasi-2D nature.
The comparison with the band-structure calculation results show that the \Uf bands are less dispersive and featureless than in the calculation, indicating that they are strongly renormalized because of the electron correlation effect.
The topology of the Fermi surface is similar to that of the band structure calculation, although the \Uf bands are strongly renormalized near \EF.
There is no Fermi surface around the \pnt{X} point, and previously reported Fermi surface at the \pnt{X} point  by low-$h \nu$ ARPES experiments is assigned as the surface electronic structure.

\ack
The experiment was performed under Proposal Nos 2016A3811, 2016B3811, 2017A3811, 2017B3811 and 2018A3811 at SPring-8 BL23SU.
The present work was financially supported by JSPS KAKENHI Grant Numbers 26400374 and 16H01084 (J-Physics), and 18K03553.
The authors acknowledge Dr. Miyawaki of Univ. Tokyo for providing his data-acquisition program for SES-2002 electron analyzer.

\section*{References}

\bibliographystyle{iopart-num}
\bibliography{URu2Si2_3D}

\end{document}